\RequirePackage{lineno}

\documentclass[aps,prb,twocolumn,groupedaddress]{revtex4}

\usepackage{amsmath,amssymb}
\usepackage{graphicx}
\usepackage{hyperref}
\modulolinenumbers[5]

\begin{document}

%\begin{frontmatter}

\title{Macroscopic-ranged proximity effect in graphite}

\author{Bruno Cury Camargo}
\email[b.c\_camargo@yahoo.com.br]{}
\affiliation{Institute of Physics, Polish Academy of Sciences, Aleja Lotnikow 32/46, PL-02-668 Warsaw, Poland.}
\author{Piotr Gier{\l}owski}
\affiliation{Institute of Physics, Polish Academy of Sciences, Aleja Lotnikow 32/46, PL-02-668 Warsaw, Poland.}
%\affiliation{Laboratoire National des Champs Magnetiques Intenses, CNRS-INSA-UJF-UPS, UPR3228; 143 avenue de Rangueil, F-31400 Toulouse, France.}
\author{Marek Kuzmiak}
\affiliation{Centre of Low Temperature Physics, Institute of Experimental Physics, Slovak Academy of Sciences, 04001 Ko\v{s}ice, Slovakia}
\affiliation{Faculty of Electrical Engineering and Informatics, Technical University, 04001 Ko\v{s}ice, Slovakia}
\author{Ramon Ferreira de Jesus}
\affiliation{Instituto Federal do Rio Grande do Sul, Campus Vacaria, 952000-000 Vacaria, RS, Brazil.}
%\author{Paulo Purreur}
%\affiliation{Instituto de Fisica, Universidade do Rio Grande do Sul, 91501-970 Porto Alegre, Brasil.}
\author{Oleksandr Onufriienko}
\affiliation{Centre of Low Temperature Physics, Institute of Experimental Physics, Slovak Academy of Sciences, 04001 Ko\v{s}ice, Slovakia}
\author{Pavol Szab\'{o}}
\affiliation{Centre of Low Temperature Physics, Institute of Experimental Physics, Slovak Academy of Sciences, 04001 Ko\v{s}ice, Slovakia}
\author{Yakov Kopelevich}
\affiliation{Instituto de Fisica Gleb Wattaghin, R. Sergio Buarque de Holanda 777, 13083-859 Campinas, Brasil.}

\date{\today}

\begin{abstract}
We report the induction of proximity-induced superconducting features over macroscopic lengths in highly oriented pyrolitic graphite (HOPG). The phenomenon is triggered when electrical currents are injected in the material through superconducting electrodes, few millimeters apart from each other. Such large range is anomalous, as proximity-induced features in normal conductors hardly surpass few micrometers. The results can be explained  as due presence of pre-existing superconductivity in graphite on small, localized regions.
 %  universal surface states expected in Dirac semimetals, and compare graphite with other Dirac and conventional materials
\end{abstract}

%\linenumbers
%\keywords{}

\maketitle
%\linenumbers

\section{Introduction}

When superconducting (S) and normal (N) materials are brought together, Cooper pairs from the S region drift into N. This is known as the superconducting proximity effect (PE), which causes a region of N close to the S-N interface to present superconductivity (SC). The effect occurs over distances of the order $\xi_N$, the normal coherence length in N. This value depends on the ratio between the superconducting coherence length in S ($\xi_s$) and the mean-free path of carriers in N ($l$). For the limiting cases $l \gg \xi_s$ and $l \ll \xi_s$ – the clean and dirty limits –, $\xi_N$ assumes the form $\xi_N = \hbar v_f / 2\pi k_B T$ and $\xi_N = \sqrt{\hbar v_F l / 6\pi k_B T}$, respectively, \cite{DeGennes1964} with $v_F$ the Fermi velocity in N . 

Although usually confined to regions tenths or hundredths of nanometers near the S-N interface, in some cases, the PE can occur over several thousand times the length $\xi_N$. This is observed in selected S-N-S systems, and is related mostly to the properties of the N material. In underdoped cuprates, for example, such a behavior has been tentatively attributed \cite{Bozovic2004, Decca2000,  Marchand2008} to the occurrence of superconducting fluctuations above $T_c$. In  clean transition-edge sensors, on the other hand, the survival of the PE over scales thousands of times higher than $\xi_N$ is thought to happen due to the presence of nonequilibrium superconductivity \cite{Sadleir2011, Sadleir2010}. In addition, geometrical quantization of superconducting excitations in clean N materials can protect supercurrents over length scales much above $\xi_N$ (see, e.g. \cite{Klapwijk1982, Kulik1970}).

One possibility to obtain the PE over macroscopic scales, hence, is to study the properties of S-N-S sandwiches with clean N materials possessing both large $\xi_N$ and indications of superconducting fluctuations. A promising candidate satisfying both conditions is graphite, which can be described as a quasi-compensated layered semimetal. In this material \cite{Zhou2006, Garcia2008, Camargo2016}, $\rho_a/\rho_c < 10^{-4}$, $l \lesssim 10$ $\mu$m and $v_F \approx 10^6$ m/s. These values lead to large estimated normal coherence lengths at low temperatures $\xi_N (T=4\text{ K}) = \hbar v_F / 2\pi k_B T \approx 2$ $\mu$m. It also presents various indications of SC, reported across the literature. Among them, are the presence of switching features in magnetoresistance akin to Josephson junction arrays \cite{Antonowicz1974, Kopelevich2003, Lebedev2014, Ballestar2015}, signatures of a  Bose-metal phase \cite{Kopelevich2006, Kopelevich2000} (also seen in Bi) and the existence of superconducting-like magnetization hysteresis loops even at high temperatures \cite{Kopelevich2000}. In addition, very recently, percolative ($R = 0$) superconductivity has been measured in twisted bi- \cite{Cao2018} and multi-layer \cite{Liu2019} graphenes.
%In addition, this material has been reported to house robust surface states susceptible to superconducting instabilities, in the form of flat-band electronic dispersions \cite{Fujita1996, Nakada1996, Shtanko2018}.
%More recently, however, it has been observed that twisted multilayer graphene samples present intrinsic superconductivity, apparently triggered for specific twisting angles between subsequent graphene layers \cite{}. Such twisting is a long-known feature in, among others, highly oriented graphite. In such material, twisting between multigraphene regions occur randomly accross the sample structure, potentially generating a number of regions with superconducting pairing potential.

 Given these signatures, it is conceivable the superconducting PE in graphite can have a far longer reach than $\xi_N \approx 1$ $\mu$m, possibly reaching macroscopic scales. So far, however, no such a phenomenon has been observed. Instead, most available reports on the literature focus on nanometer-sized devices \cite{Hayashi2008, Sato2008, Lee2013, Sato2008v2, Sangiao2017}. Here, we studied the electrical transport characteristics of millimetric samples outfitted with superconducting current-injection leads. Results revealed signatures of a long-range PE in our devices,  persisting above $700$ $\mu$m ($200-300$ $\xi_N$) from the superconducting electrodes - an unusually long distance in bulk systems.

\section{Results and discussion}

\subsection{Sample preparation} \label{sec_sample}

All our samples were extracted from a commercially-available highly oriented pyrolytic graphite (HOPG) crystal with $0.3^o$  mosaicity \cite{GW_INC}. Its room-temperature in-plane resistivity was, approximately, $5$ $\mu \Omega$.cm. The devices had typical in-plane dimensions of $5$ mm $\times$ $1$ mm and thicknesses varying between $0.15$ mm and  $0.4$ mm. 

Samples were either contacted in a standard $4$-probe configuration for electrical transport measurements, or in a 6-probe Hall bar configuration for simultaneous Hall and longitudinal resistance measurements. Samples contacted in a 4-probe configuration had voltage electrodes covering the whole sample's surface width, as is schematized in the inset of Fig. \ref{fig_pristine}. This was done to mitigate possible effects of current distribution in the material. Voltage electrodes of samples contacted in the 6-probe Hall bar configuration were spread across the whole height of the sample, for the same reasons (see Fig. \ref{fig_Hall}). 

Additional devices were prepared in a 8-probe longitudinal/Hall-bar hybrid configuration. In it, electrodes were placed at the surface and edges of the samples. This geometry was realized to verify possible current distribution issues. In these devices, edge electrodes covered the whole sample height (as for devices with 6 probes), whereas top electrodes were point-like (see Fig. \ref{fig_8p}).

\subsection{Control samples and the superconducting alloy} 

Control samples were characterized in the interval $2\text{ K} \leq \text{T} \leq 10 \text{ K}$ with all electrodes made of silver paste. Measurements revealed a smooth metallic-like behavior, typical of well-graphitized HOPG \cite{Kelly_book, Du2005, Hamada1981}. Results are presented in Fig. \ref{fig_pristine}. 

Subsequent samples were contacted as indicated in the cartoon of Fig. \ref{fig_pristine}, with two outermost electrodes composed of a superconductor (SC) and the remaining ones of silver paste. The distance between adjacent electrodes was approximately $0.7$ mm to $1$ mm. We refer to their contacts by the numbers shown in Fig. \ref{fig_pristine}.  

In total, ten samples were studied (labeled $GS1$ to $GS10$). The electrical resistivity of our devices was probed with DC and low-frequency AC measurements (up to $f \approx 5$ Hz), which yielded the same results. Experiments were simultaneously performed in two-and four- probe configurations. Two-probe measurements ($2p$) were achieved by applying current and measuring voltage between superconducting electrodes $1$ and $4$. Experiments in the four-probe configuration ($4p$) were performed by applying electrical current between (superconducting) contacts 1 and 4 and measuring the potential between the (normal) probes $2$ and $3$. 

The superconductor chosen for the current electrodes was an alloy of In/Pb ($50\%$ in volume each) with critical temperature $T_c \approx 6.8$ K and critical magnetic field $B_c(0) < 1.8$ T (see the suppl. material for details). The alloy was connected to the edges of our samples with a soldering iron at $300^o$ C. Thin copper wires attached the sample to the instrumentation.

%\section{Results and discussion}

\subsection{Transport measurements} \label{sec_transport}

\begin{figure}[h]
\begin{center}
\includegraphics[width = 8cm]{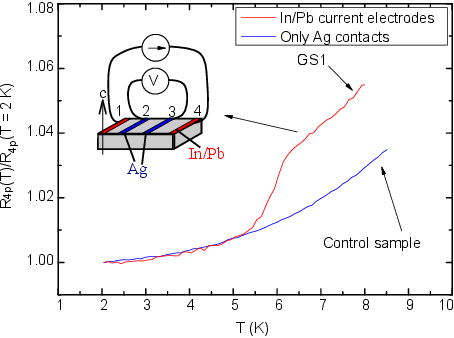}
\caption{Normalized four-probe $R_{4p}(T)$ curves for the control (pristine) graphite sample (blue curve) and for the device $GS1$ (red curve) at zero magnetic field. The normalization factors were $R_{4p}(T = 2\text{ K}) = 3.66$ m$\Omega$ for the control  and $R_{4p}(T = 2\text{ K}) = 4.34$ m$\Omega$ for $GS1$. The inset shows a sample schematic. Numbers $1-4$ identify each electrode.  The control sample was measured as indicated for $GS1$, but with all contacts made of non-superconducting silver paste.}
\label{fig_pristine}
\end{center}
\end{figure}

The main result of the present work is illustrated in Fig. \ref{fig_pristine}. In it, resistance vs. temperature plots of the sample $GS1$ are shown. Measurements were carried out prior and after connecting SC current-injection leads to the device. While no anomaly was observed for a sample with only N electrodes, the injection of electrical currents through SC contacts induced a superconducting-like transition in $4p$ measurements. This occured despite voltage probes being distant ca. $0.7$ mm - $1$ mm from the SC contacts (two orders of magnitude above $\xi_N$). Such a behavior was shared by all our devices. 

The details of the $4p$ feature were sample dependent. In eight of our devices, the transition manifested as a sharp decrease of the $4p$ resistance ($R_{4p} \equiv V_{4p}/I$, $I$ the applied current) below Tc. In one sample, it manifested as a sharp increase. One device did not present signatures of the 4p transition, within experimental uncertainty. When a decrease was observed, it accounted for, at most, $10\%$  of the sample resistance at the center of the transition. Its general behavior was similar to the one observed in  $2p$ measurements ($R_{2p} \equiv V_{2p}/I$), which are undoubtedly associated to the superconductivity at the SC electrodes. The large distances between the normal and superconducting contacts in the $4p$ configuration, however, do not allow for the same interpretation. In what follows, we consider data of the majority of devices, showing a sharp decrease in $R_{4p}(T<T_c)$. The discussion, however, can also be applied to the case when a sharp increase manifests.

\begin{figure}[h]
\begin{center}
\includegraphics[width = 8cm]{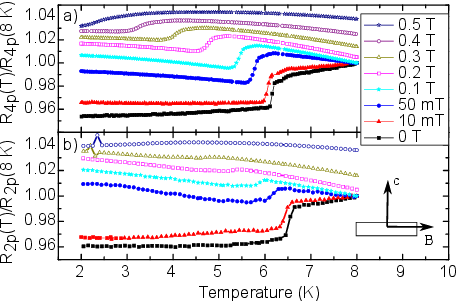}
\caption{a) $4$-probe and b) $2$-probe resistance of sample $GS2$ as a function of temperature for different magnetic fields. The resistance has been normalized by its value in $8$ K and the curves have been displaced vertically for clarity. The sample was measured as shown in the cartoon in fig. \ref{fig_pristine}, with the magnetic field applied parallel to the graphene planes.}
\label{fig_RxTGS1}
\end{center}
\end{figure}

Figure \ref{fig_RxTGS1} shows the behavior of sample $GS2$ in the presence of in-plane magnetic fields ($B\bot \text{c}$)  for the $2p$ and $4p$ configurations. This orientation was chosen to suppress the high orbital magnetoresistance of graphite, which can reach $1000$ \%  for relatively low ($\approx 0.1~T$) magnetic fields along the material's c-axis \cite{Camargo2016}. Because the phenomenon of interest was superimposed to this response, the chosen geometry ensured the best possible experimental resolution. Similar results were obtained for $B//\text{c}$ and are shown in the suppl. material.  In the sample, the electrical current leads presented a contact resistance $R_C \approx 12.5$ m$\Omega$ at $T = 2$ K, while graphite ($R_{2p} - 2R_C$) had a resistance of approx. $32.3$ m$\Omega$. Two-probe measurements revealed a clear superconducting transition with $T_c \approx 6.8$ K, accounting for up to $6\%$ of the total sample resistance at zero magnetic field. A transition was also observed in 4p measurements in the same temperature range, albeit accounting for up to $3\%$ of the measured signal.% Other samples (see the suppl. material) presented $2p$ transitions accounting up to $15\%$ of the sample resistance, and $4p$ transitions up to $5\%$.

\begin{figure}[h]
\begin{center} 
\includegraphics[width = 8cm]{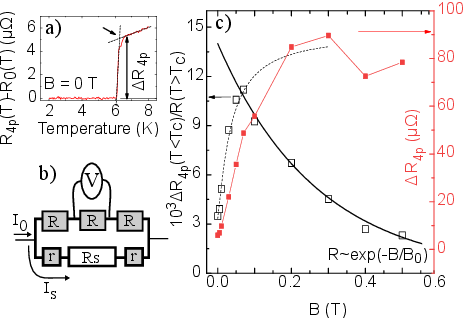}
\caption{a) Example of how the transition amplitude $\Delta R_{4p}$ was determined. The y-axis corresponds to $R_{4p}$ of sample $GS2$ at zero magnetic field after subtraction of a polynomial resistance background $R_0(T)$. $\Delta R_{4p}$ is chosen at the point indicated by an arrow in the figure. This value corresponds to the intersection between the maximum slope during the transition and a linear extrapolation of the $R_{4p}$ behavior above $T_c$. b) Cartoon representing the proposed equivalent circuit for the sample. Below $T_c$, a low resistance channel $R_s$ forms, which carries a fraction $I_s$ of the total applied current $I_0$. This results in a step of amplitude $\Delta R_{4p} = \Delta V / I_0  = R(T_c)\times I_s/I_0$, where $R(T_c)$ is the resistance of a pristine sample at $T_c$. c) Normalized (open symbols, left axis) and non-normalized (right axis, closed symbols) amplitude of the $4p$ transition as a function of the magnetic field in the device $GS2$. The dashed and solid lines are functions of the type $y = \alpha R/(r+R)$ and $y=\beta exp(-B/B_0)$, respectively. On them, $R = (35 + 6.7 \times 10^4 B^{1.2})$ m$\Omega$, $r=50$ m$\Omega$ and $B_0 = 0.13$ T.}
\label{fig_ampl_vs_B}
\end{center}
\end{figure}

The amplitude of the $4p$ transition, defined as $\Delta R_{4p} \equiv |R_{4p}(T\lesssim T_c)-R_{4p}(T\gtrsim T_c)|$ (see fig. \ref{fig_ampl_vs_B}), did not vary monotonically as a function of $B$. Instead, it increased with the applied magnetic field, reaching saturation above 0.1~T (see Fig. \ref{fig_ampl_vs_B}). This behavior can be understood by considering the creation of a low-resistance channel below $T_c$, which operates in parallel with graphite. The equivalent circuit is represented as a cartoon in Fig. \ref{fig_ampl_vs_B}. In it, the low resistance channel (denoted by $R_s$) acts as a shunt resistor, which carries a fraction $I_s$ of the total electrical current $I_0$ applied to the system. Assuming that $R_{4p}$ senses mostly the dissipative channel of graphite, the \textit{measured} resistance  below $T_c$ can be expressed as a function of $I_s$ by %drop below $\text{T}_\text{c}$ becomes proportional to the product between the shunted current and the resistance $R$ of pristine HOPG at a given temperature:
\begin{equation}
R_{4p}(T<T_c) \equiv \frac{V_{4p}(T<T_c)}{I_0} = \frac{R(T)\times (I_0-I_s)}{I_0},
\label{eq_res_drop}
\end{equation}
where $V_{4p}(T<T_c) = R(T)\times(I_s-I_0)$ corresponds to the actual \textit{measured} voltage drop between normal electrodes on the sample, and $R(T)$ is the resistance of pristine graphite with normal electrodes.

Above $T_c$, $R_{4p}(T)$ should match the resistance of the pristine sample, as the low resistance channel ceases to exist ($I_s = 0$). In this case $R_{4p}(T>T_c) = R(T)\times I_0$ and, through eq. \ref{eq_res_drop}, we obtain a simple expression for the amplitude of the $4p$ transition 
\begin{equation}
\Delta R_{4p} \equiv R_{4p}(T\lesssim T_c)-R_{4p}(T\gtrsim T_c) \approx R(T_c)\times \frac{I_s}{I_0}.
\label{eq_transition_ampl}
\end{equation}

Considering the weak slope on the $R(T)$ behavior of pristine graphite (see Fig. \ref{fig_pristine}), $R(T_c)$ is approximated by $R(T_c) \approx R_{4p}(T\gtrsim T_c) $. The expression \ref{eq_transition_ampl} can then be normalized, resulting in 
\begin{equation}
\frac{\Delta R_{4p}}{R_{4p}(T\gtrsim T_c)}  \approx  \frac{I_s}{I_0} \propto I_s
\label{eq_res_drop_2}
\end{equation}
and providing a link between $\Delta R_{4p}$ and $I_s$.%for a method to analyze the non-monotonic behavior shown in fig. \ref{fig_ampl_vs_B}c.
%Changes in the transition amplitude can, therefore, be directly associated with small variations of $I_s$.

We first consider the behavior of $\Delta R_{4p}/R$  at weak magnetic fields $B<0.1$ T (see fig. \ref{fig_ampl_vs_B}c). In this field range, our data can be described with the circuit shown in Fig. \ref{fig_ampl_vs_B}b by assuming a low resistive channel $R_s \ll R$  weakly affected by small values of $B$, and $I_s$ being limited by some interface resistance $r$. In this case $I_s = \Delta R_{4p}/[R_{4p}(T>T_c)]\approx 3R/(3R+2r)$.  The expression describes well the experimental data if we take $r$ constant and $R\propto B^{1.2}$, which is the typical magnetoresistance of graphite \cite{Camargo2016}. %The hidden assumption in eqs. \ref{eq_res_drop} - \ref{eq_res_drop_2} is that $I_s(T<T_c)$ does not change with T. This is consistent with the observation that $R_4p(T<T_c)$ of GS samples below $T_c$ does not diverge from the $R(T<7K)$ behavior of pristine graphite (see fig. \ref{fig_pristine}) and supports the presence of localized superconducting domains in the material, as will be discussed later on.%The latter  between the low resistance channel and the superconducting electrodes.

At higher magnetic fields, on the other hand, the decay observed for $\Delta R_{4p}/R \propto I_s$ indicates a progressive suppression of the low resistance channel by $B$. In this field range, $I_s$ has the functional form $I_s\propto exp(-B/B_0)$. This is the same dependency expected for the critical current density through a thick superconductor-metal-superconductor Josephson junction with increasing magnetic fields \cite{Dobro1993}, further suggesting the $4p$ transition as a superconducting proximity-induced feature in our system. The extracted parameter $B_0$ was $B_0 \approx 0.13$ T. This value practically coincides with the critical magnetic field $B_{crit} \approx 0.11$ T introduced in Ref.\cite{Kopelevich2003}, which corresponds to the magnetic field necessary to suppress the bosonic character of carriers in graphite. In other words, magnetic fields in this scale have been proposed to destroy electron-electron pairing in the system \cite{Kopelevich2003}.%In it, the magnetic-field-induced-metal-insulator transition in pristine graphite in terms of a Bose-metal-to-insulator quantum phase transition.
%This parameter $B_{crit}$ and corresponds to the magnetic field necessary to suppress the Bose-metal phase in the system. In other words, it destroys the bosonic character of carriers in the material, providing a characteristic magnetic field scale necessary to destroy localized superconducting domains in the system.

Although the amplitude of the $4p$ anomaly did not change monotonically with T, the temperature in which it occurred was displaced by the presence of magnetic fields according to the empirical expression
\begin{equation}
B_c(T) = B_c(0) \left(1-\frac{T_c}{T_{c0}} \right),
\label{eq_phase_diagram}
\end{equation}
which correlates the transition temperature $T_c$ with the applied field $B_c$. In the equation, $B_c(0)$ corresponds to the critical magnetic field at $T = 0$ K, and $T_{c0}$ the critical temperature at zero magnetic field. %As the magnetic field was increased, the transition progressively blended with the background signal, which developed an insulating-like behavior with saturation for $\text{T} < 5$ K. Such insulating phase is typical of graphite, and was previously reported by different authors \cite{Kopelevich2006, Kopelevich2000, Du2005, Kempa2002, Edwards1998}.

\begin{figure}[h]
\begin{center}
\includegraphics[width = 8cm]{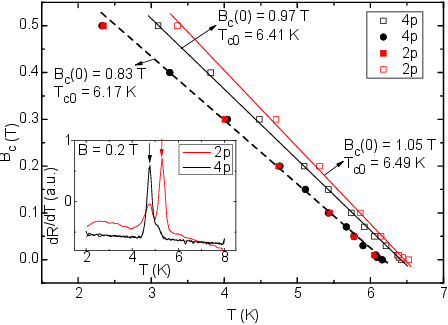}
\caption{$B_c(T)$ vs. $T$ diagram of sample $GS1$ for the $2p$ (red symbols) and $4p$ (black symbols) transitions. Open and closed symbols refer to subsequent transitions from the same R(T) curve, happening at higher and lower temperatures, respectively. The lines follow from eq. \ref{eq_phase_diagram} using the parameters $B_c(0)$ and $T_{c0}$ shown in the figure. The inset shows the derivative of the $2p$ and $4p$ resistance as a function of temperature. The maxima in such curves were chosen as the transition temperature.}
\label{fig_BxTdiagram}
\end{center}
\end{figure}

The $B_c(T)$ diagrams for sample $GS2$ are shown in Fig. \ref{fig_BxTdiagram}. They revealed that the $4p$ feature happened at lower temperatures and magnetic fields in relation to the $2p$ transition. This result suggests the 4p transition as a consequence of the $2p$ one. Indeed, both $2p$ and $4p$ measurements revealed two transitions occurring in close proximity, which can be linked to the existence of two phases in the SC electrodes, with similar $T_c$'s (see the suppl. material). The persistence these features in $4p$ measurements indicate that the suppression of the $4p$ transition by magnetic fields is closely related to the suppression of superconductivity at the current electrodes \cite{Han2014}.% In particular, the transition with lowest $\text{T}_\text{c}$ is clearly observable in $2p$ and $4p$ measurements, although presenting a higher relative intensity on the latter. This can be explained by considering such transition as the one taking place in graphite (which would also be probed in $2p$ measurements, albeit with smaller relative intensity).
 This is also reinforced by the fact that the $B_c$ vs. $T$ diagram for the $4p$ transition remained unchanged after modifying the magnetic field orientation relative to the sample's c-axis (see the suppl. material). Such a result strengthens the hypothesis of superconducting electrodes as the objects inducing the phenomenon at hand. It also shows that the magnetic flux through the regions of HOPG responsible for the $4p$ transition does not depend on the magnetic field orientation. In a highly anisotropic system such as graphite, the latter can be explained by a low resistance channel confined to low-dimensional (quasi-1D or quasi-0D) sites/structures within the material.

\begin{figure}[h]
\begin{center}
\includegraphics[width = 8cm]{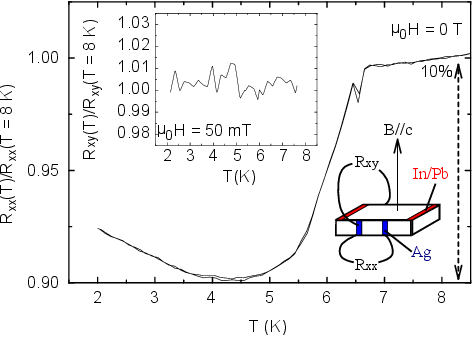}
\caption{Longitudinal resistance normalized by its value at $T=8$~K for sample $GS5$, measured at zero magnetic field. The sample was contacted at the edges in a 6-probe Hall configuration, as shown in the cartoon. The inset shows the Hall resistance extracted at $B = 50$~mT, also normalized by its value at 8~K. Note the lack of a transition in the Hall component of resistance.}
\label{fig_Hall}
\end{center}
\end{figure}

The existence of these regions/structures would also explain the behavior of $R_{4p}(T)$ in our devices, which did not show a continuous drop below Tc when compared with pristine samples (see Fig. \ref{fig_pristine}). Such a result is at odds with the conventional superconducting PE. In it, the $4p$ resistance should change linearly with $\xi_N$, roughly\cite{DeGennes1964, Kompaniiets2014} as $R_{4p}(T) \propto R_0 (1-\kappa \xi_N)$, $\kappa \xi_N \propto 1/T$. Instead, our $R(T)$ results suggest that an eventual proximity-induced state seen in $4p$ must be confined to a fixed fraction of the sample volume, as also inferred from measurements performed at $B \bot c$ and $B//c$. This hypothesis is further supported by measurements in samples contacted in the 6p-Hall configuration, see fig. \ref{fig_Hall}. On them, the resistance drop below $T_c$, observed in the longitudinal resistance, was larger than on other configurations (ranging about 10\% of the total measured signal). Meanwhile, a negligible change in the Hall effect was observed in the same temperature range. Such results are consistent with the induction of the phenomenon at small, localized puddles on the sample, which usually do not contribute to the material's Hall conductivity. % By construction, the magnetic flux through such regions should be small and independent on the magnetic field orientation.% It also suggests that the magnetic flux through the regions of HOPG responsible for the $4p$ transition does not depend on the magnetic field orientation.

\begin{figure}[h]
\begin{center} 
\includegraphics[width = 8cm]{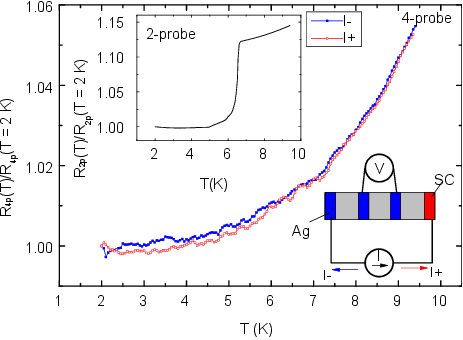}
\caption{Normalized resistance vs. temperature for a graphite sample contacted in an asymmetric configuration (only one superconducting electrode). Curves in the main panel show DC measurements, performed using different electrical current polarities. The cartoon on the bottom right illustrates how the experiment was performed. The inset shows data obtained in the $2p$ configuration.}
\label{fig_null_results}
\end{center}
\end{figure}

The absence of the \textit{conventional} superconducting PE in our samples is further evidenced by experiments in asymmetric devices with a single superconducting electrode. Measurements in this configuration did not present any anomaly (see Fig. \ref{fig_null_results}). These results also weight against experimental artifacts due to the distribution of electrical currents. Had this been the case here, the presence of a single superconducting electrode should suffice to prompt changes near at least one of the voltage probes, therefore, triggering the same effect seen in $4p$.

\subsection{Surface characterization}

We proceed to verify if our observations could be an artifact caused by the diffusion of superconducting material throughout graphite, originating from the current electrodes, or due to the dispersion of a ``cloud'' of superconducting particles across the sample surface during soldering. 

\begin{figure}[h]
\begin{center} 
\includegraphics[width = 8cm]{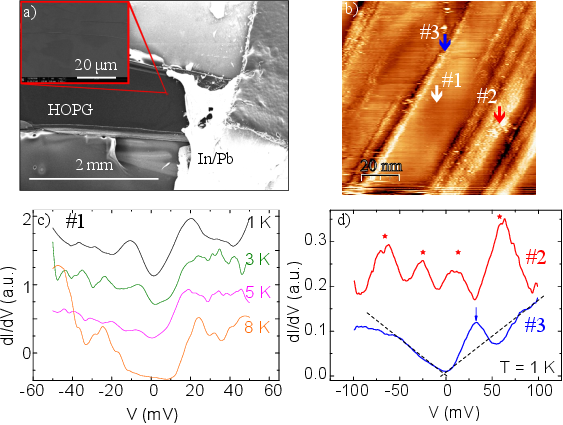}
\caption{a) SEM micrograph of a HOPG sample with a soldered In/Pb electrode (white area). The magnification corresponds to a 50 $\mu$m $\times$ 50 $\mu$m region c.a. 200 $\mu$m distant from the superconducting electrode. No scattered material is observed. b)  Typical topography of the region between electrodes for the graphite samples. The topography was measured at $T = 1$~K in an STM microscope. The arrows \#1, \#2 and \#3 point the regions where the curves in c) and d) were obtained. c) Typical non-featureless differential conductance curves obtained in flat regions of STM scans. Curves have been displaced vertically for clarity. A gap-like structure is observed at low temperatures, which is not suppressed as $T$ increases.  d) Typical differential conductance features obtained at buckled (\#2) and step regions (\#3) at $T=1$~K. The peaks in \#2, indicated by stars, are nearly periodic in V and can be associated to a pseudomagnetic field of 2.7~T assuming an electronic effective mass for graphite $m\approx 0.05m_e$, $m_e$ the free electron mass \cite{Chung2002}. The straight lines in \#3 are a guide to the eye, and indicate a graphene-like linear energy dispersion near the Fermi level \cite{Andrei_2012}. The peak pointed by an arrow occurs around 30~mV and is associated to edge states of zig-zag terminated graphene flakes \cite{Kobayashi2005}.}
\label{fig_STM}
\end{center}
\end{figure}

Both possibilities are unlikely due to the characteristic time lengths involved in the soldering process (few seconds) and to the fact that no flux is used (hence, no ``particle cloud'' should form). For confirmation, scanning electron microscopy (SEM) measurements and extensive low-temperature scanning tunneling spectroscopy (STS) at different points across sample surfaces were carried out. Superconducting dust splattered during soldering should be observable in SEM images as small particles. Diffusion of superconducting material throughout graphite should manifest as superconducting gaps of amplitude $\approx 1$ meV (the same one of lead) on scanning tunneling spectra measured over large sections of HOPG. Neither of these features were observed. In particular, scanning tunneling microscopy (STM) measurements in HOPG did not show qualitative differences above and below its electrode's $T_c$. Combined, these observations categorically discard In/Pb diffusion or localized foreign superconducting islands as the source of our observations. Examples of SEM images of the sample surface, together with typical STM spectra are shown in fig. \ref{fig_STM}.

Despite not showing signatures of superconductivity, STS did reveal features similar to those of graphite/graphene. These include occasional regions with graphene-like spectra, periodic peaks on the density of states as a function of bias voltage and small pseudo-gap-like features, with energies in the range of 10 meV (see Fig. \ref{fig_STM}c,d). The latter have been previously observed and were associated with van-Hove singularities caused by the twisting of adjacent graphene layers. The obtained gap width (10 meV) corresponds to a rotation between layers \cite{Andrei_2012} around $1.13^o$. This value is close to the ``magic angles'' for which bilayer graphene has been recently reported to present superconductivity \cite{Yankowitz2018}. Such a result provides a candidate for the localized regions subject to induced superconductivity in our samples.

Additionally, STM measurements made in the contact regime (i.e., the probing tip was touching graphite's surface) presented zero bias conductance peaks in some regions of the sample surface. They manifested as a nearly two-fold enhancement of the local differential conductance in graphite. Such a feature was well-described within the Blonder-Tinkham-Klapwijk model for a direct ballistic N-S point contact junction \cite{Plecenik1994}. This feature, which is shown in the supplementary material, will be discussed in more detail elsewhere. Such results suggests the possibility of Andreev reflections of quasiparticles in a ballistic microstriction between a normal metal and a superconducting region, occurring during measurements in different parts of the sample.%, thus further indicating the existence of superconducting islands scattered across graphite. 

\subsection{Current distribution}

In anisotropic materials (such as graphite), contact placement can aggravate current distribution issues. These, in turn, can cause artifacts that compromise measurements. In this work, we minimize possible artifacts arising from current distribution by choosing different contact configurations/geometries, by measuring several devices, and by probing different samples.

As explained in Sec. \ref{sec_sample}, samples were contacted with three different probe configurations. Two of those are represented in figs. \ref{fig_pristine} and \ref{fig_Hall}. Namely, samples with 4 probes were measured as schematized in the insert of fig. \ref{fig_pristine}, with electrodes spanning the whole width of the sample. Conversely, samples contacted in the 6-probe Hall configuration (Fig. \ref{fig_Hall}) had voltage electrodes placed only at the sample's edges, covering its entire height. Yet, in both cases, a clear transition to a lower resistance state was observed below $T_c$ (see Figs. \ref{fig_pristine}, \ref{fig_RxTGS1} and \ref{fig_Hall}). 

The same qualitative behavior in both 4-probe and 6-probe-Hall geometries indicate that our samples are not subject to current distribution issues. Otherwise, qualitatively different results would be expected, hinging on contact positioning. This argument is supported by measurements performed in samples containing a single superconducting electrode. Results for one of these devices are shown in Fig. \ref{fig_null_results}. All samples measured in this configuration (three in total, see the Suppl. material) did not present transitions below $T_c$. Had the origin of the $4p$ transition been associated with current distribution caused by the SC electrodes, it should also manifest on those samples.

\begin{figure}[h]
\begin{center} 
\includegraphics[width = 8cm]{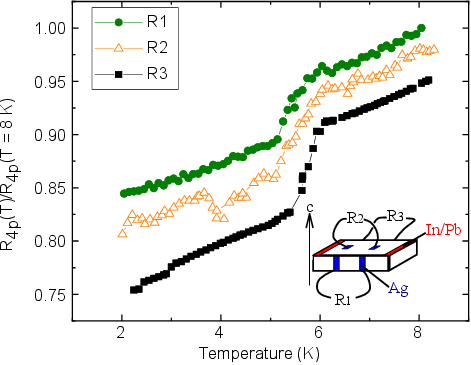}
\caption{Normalized $R_{4p}(T)/R_{4p}(T = 2 K)$ measurements for a sample prepared with 8 electrodes (GS10), as represented in the inset of the figure. The current electrodes were made of a Pb/In alloy, whereas all voltage electrodes were composed by Ag paste. $R1$, $R2$ and $R3$  correspond to the respective electrode pairs shown in the inset. The normalizing factors were $R_{4p}(T = 2K) = 7\mu \Omega$, 3 $\mu Omega$ and 13 $\mu \Omega$  for R1, R2 and R3, respectively. R2 and R3 were measured along the sample edges, whereas R1 was measured at the sample top.}
\label{fig_8p}
\end{center}
\end{figure}

Despite all our samples being prepared using the same raw materials, measurements performed on the 4-probe and 6-probe-Hall configurations, as shown in Figs. \ref{fig_RxTGS1} and \ref{fig_Hall}, were realized on different devices. To account for this, a sample was prepared in an 8-probe hybrid configuration. This geometry was a combination of the 4-probe and 6-probe Hall geometries, thus allowing the simultaneous measurement of the longitudinal $4p$ sample resistance at the top and along the edges of the sample. A schematic is shown on the inset of Fig. \ref{fig_8p}. Due to conservation of charge, current distribution issues on this device would require qualitatively different behaviors to be observed on different electrodes. Instead, simmmilar measurements were obtained for both top and edge electrodes, corroborating the results presented in figs. \ref{fig_RxTGS1} and \ref{fig_Hall}. Such a result further discards current inhomogeneities as the source of the $4p$ transition reported here. 

Finally, we verified possible instabilities of the instrumentation used during the experiments. For this, a piece of copper was probed under the same conditions as used for graphite samples. Results did not reveal any transitions (see the supplementary material), discarding instrumentation artifacts from current sources, amplifiers and voltmeters as the source of our observations in graphite.

\subsection{Discussion}

The overall behavior of the $4p$ transition triggered in graphite supports the presence of induced superconducting features in our devices in ranges above those of the conventional superconducting PE. Its characteristics, however, do not point towards a bulk-related phenomenon.%; magnetotransport and Hall measurements indicating an effect confined to a small volumetric fraction of the sample, in regions of reduced dimensionality.% and the weak dependence on magnetic fields suggesting

As discussed in section \ref{sec_transport}, the properties of our samples can be described by the existence of two independent transport channels acting in parallel: a high- and a low-resistance one. The high resistance channel seems to be related to transport through pristine graphite, whereas the low resistance channel is unambiguosly associated to the presence of superconducting electrodes in the device. The fact that Hall measurements remain unchanged above and below $T_c$, whereas the sample's longitudinal resistance shows variations as large as 10\%, strongly suggests that transport through the low resistance channel occurs along small localized grains in graphite, which should not contribute to its Hall resistance.% Meanwhile, transport by normal carriers happen through the normal matrix. Hence, we can consider our sample as an array of mesoscopic superconducting islands embedded in a graphitic matrix, similarly to the systems reported in refs. \cite{Han2014, Allain2012, Kessler2010}. 

Suitably, the occurrence of the \textit{conventional} superconducting PE does not justify the $4p$ transition. As previously presented, the lack of features in samples with a single SC electrode, the null dependence on T, the increasing transition amplitude with magnetic fields, and the long range of the phenomenon (up to $700$ $\mu$m - $1$ mm distant from the superconducting electrodes, whereas $\xi_N\approx 1 - 2$ $\mu$m \cite{DeGennes1964, Zhou2006, Garcia2008, Camargo2016, Pippard1952, Beletskii1975, Gantmakher1971}) all attest against such a hypothesis. 

Instead, all our results can be accounted for by considering the pre-existence of mesoscopic, superconducting-prone islands in graphite. This hipothesys is consistent with previous reports in the literature, which indicate localized superconducting domains in the material \cite{Kopelevich2006, Antonowicz1974, Lebedev2014, Ballestar2015, Precker2016}. In this context, the presence of superconducting leads can act as a trigger for global coherent transport along such a pre-existing channel. In addition to overall sample behavior,  superconducting puddles can be justified by point contact measurements performed during our STM study.  These provided strong indications of Andreev reflections across graphite's surface, signalizing the existence of superconducting regions in bulk HOPG with dimensions above the superconducting coherence length \cite{Deutscher1999}. 

Pristine samples, however, showed no superconducting-like transitions, in agreement with most reports to date \cite{Camargo2016, Sato2008, Kopelevich2000, Soule1958, Kempa2002}. Such  an observation requires that the transfer of charge from bulk graphite to the hypothesized superconducting regions must be forbidden under normal conditions. This can be justified self-consistently when considering the superconducting islands in graphite as objects with reduced dimensions that are embedded in a quasi-$2$D electron gas with low conductance. This hypothesis has been suggested on previous STM, magnetization and transport studies in different types of pristine graphite \cite{Kopelevich2006, Antonowicz1974, Lebedev2014, Ballestar2015, Du2005}, as well as inferred from our measurements.

Indeed, considering the typical $4p$ sample resistivity around $\sigma \approx 5$~$\mu \Omega$.cm (see Fig. \ref{fig_pristine}) and assuming a homogeneous current distribution across the sample volume, our devices show a conductance per graphene layer of the order $\sigma_L = \sigma c_0/t \approx 4.6 \times 10^{-4}$  S  $\approx 6 G_0$. In it, $c_0=0.335$~nm is the interplane distancing in graphite, $t\approx 0.2$~mm the sample thickness and $G_0=2e^2/h$. In this case, the system could be roughly described as an array of superconducting islands at $T \ll T_c$, embedded in a quasi-$2$D electron gas (2DEG) with local conductance close to the conductance quantum. This situation is similar to a graphene film decorated with superconductors, tuned near the charge neutrality point \cite{Han2014}. Under such circumstances, the Coulomb blockade impeding the introduction of carriers into the superconducting islands is expected to decrease with the conductivity of the 2DEG, following $exp(-\pi^2 G_D/8)$, $G_D$ the conductance of the metallic matrix \cite{Feigelman2001, Feigelman2002}. Such an enhanced Coulomb blockade at low conductances (approaching a few kilo Ohms) can lead to a weak charge quantization in the superconducting grains. This forbids charge transfer to/from the superconducting regions. The phenomenon effectively disables the superconducting channel in the material by fixing the number of cooper pairs in the system ($\Delta N = 0$), which results in large phase fluctuations destroying the macroscopic superconducting order \cite{Feigelman2001}.

However, the introduction of superconducting leads in the sample acts as a reservoir of Cooper pairs, bypassing the weak quantization constraint and effectively delocalizing carriers ($\Delta N \neq 0$). This re-enables transport by this network of superconducting islands - thus resulting in the observed behavior in $4p$ measurements.

Our interpretation is also consistent with previous analysis of the magnetic-field-induced metal-insulator transition taking place in HOPG, which indicated intrinsic 2e carriers in the material \cite{Kopelevich2003}. In particular, the characteristic magnetic field $B_0\approx 0.13$~T related to the suppression of the $4p$ transition (fig. \ref{fig_ampl_vs_B}), is very close to the critical magnetic field $B_{crit} \approx 0.11$ T  associated with the suppression of the bosonic character of carriers in the system \cite{Kopelevich2003}. The presence of paired electrons in the absence of a coherent superconducting phase entails the existence of localized superconducting puddles. Hence, if the $4p$ transition is to be related to small superconducting-prone regions, its suppression should occur in tandem with the suppression of the Bose-metal phase discussed for graphite \cite{Kopelevich2003}. This is indeed suggested by the similarity between $B_0$ obtained here and the parameter $B_{crit}$ of ref. \onlinecite{Kopelevich2003}.

%such simmilarity between $B_0$ and bose-metal-like phase is associated with localized cooper pairs in the system, our hiporthesys 
%This suggests that pairing in graphite might be held accountable both phenomena might be related, the presence of $$See fig. \ref{fig_ampl_vs_B} above and the discussion associated.

%Our interpretation is also consistent with previous analysis of the magnetic-field-induced metal-insulator transition taking place in different types of HOPG, which indicate intrinsic 2e carriers in the material. Namely, the suppression of the $4p$ transition reported here seems to occur in tandem with the suppression of the Bose-metal phase discussed for graphite in reference \cite{Kopelevich2003}. This is suggested by the similarity between $B_0$ obtained here and the parameter $B_{crit}$ of ref. \cite{Kopelevich2003}. The latter is associated with the suppression of the bosonic character of carriers in the system. These can be understood as paired electrons in the absence of a coherent superconducting phase, which  entails the existence of localized superconducting puddles.

%The proposed hypothesis also accounts for the qualitatively different $4p$ behaviors observed in different samples (sharp increases or decreases of the $4p$ sample resistance below $\text{T}_\text{c}$, see Figs. \ref{fig_pristine} and the supplementary information), as well as for 
In the context of thick S-N-S junctions, the characteristic magnetic field $B_0\approx 0.13$ T (see fig. \ref{fig_ampl_vs_B} and the associated discussion) can  be further related to a distance $L$ between superconducting regions within graphite. This distance can be estimated as \cite{Dobro1993} $L =2\sqrt{h/(2\pi e B_0)}\approx 100$ nm. Such a value is justified by considering that low magnetic fields (prior to the suppression of superconductivity at the electrodes) disrupt the interaction between superconducting grains within graphite, rather than a macroscopic weak link between the two superconducting electrodes. Random in nature, such grains are to present a sample-dependent distribution. Hinging on their size and coupling, their network can act either as a low- or a high-resistance channel for carriers, akin to observations in homogeneous and highly granular superconducting thin films, respectively \cite{Goldman2003}.% In both cases, however, transport via this channel is conditioned to the injection of $2e$ bosons in graphite, which is achieved by the superconducting contacts.

Although not showing signs of a superconducting gap, STM measurements allow for a second, upper limit estimation of the average distance between such grains. A proximity-induced superconducting gap in a N layer of thickness L is expected to be at least 3 times larger than the structure's Thouless energy \cite{Sueur2008, Quaglio2012} $E_{Th}=\hbar D/L^2$.  In it, $D$ is the electronic diffusion coefficient on N. For graphite with electronic mobility \cite{Camargo2016} $\mu \approx 10^5 \text{ cm}^2/\text{V.s}$, $D \approx 100 \text{ cm}^2$/s at $T = 1$~K. Assuming $E_{Th} \lesssim 5 \times 10^{-4}$ meV in our samples (the experimental resolution) results in an estimated distance between superconducting puddles on graphite of the order $L\approx 1$ $\mu$m. In multigraphenes, proximity phenomena have been shown to surpass these distances \cite{Heersche2007}.

The remaining issue becomes, then, the identification of such regions. The absence of clear evidence of superconducting gaps in  STM scans seem to discard most of the features commonly found in the surface of graphite (e.g. wrinkles, folds, bubbles, impurities and grain boundaries) as possible candidates for hosting the proposed superconducting-prone regions. We are not able to eliminate, however, regions with gap-like structures in their STS (see fig. \ref{fig_STM}). These can be associated with the twisting of adjacent graphene layers by small angles \cite{Andrei_2012}. Such regions been demonstrated to host superconductivity at low temperatures. However, STM measurements specifically designed to probe their properties are yet to reveal clear evidence of superconducting gaps at temperatures as low as 1~K (see e.g. refs. \onlinecite{Jiang2019, Xie2019}). 

Additionally, experiments performed by us in samples contacted at their lateral edges (see fig. \ref{fig_Hall}) have shown resistance drops below $T_c$, amounting for about 10\% of the total measured signal. Transitions on this geometry were better-defined and larger than those observed on samples contacted at the top surface - figs. \ref {fig_RxTGS1} and \ref{fig_Hall}). Edge regions  are more susceptible to deformations during the sample cutting process, which generate different types of irregularities. Enhanced transitions observed along them further indicate lattice distortions as possible candidates for the phenomenon at hand. Unfortunately, due to the disordered nature of graphite's edges, we are currently unable to probe their differential conductance (as done for the remainder of the sample surface).

In short, we demonstrated the induction of a  macroscopic, long-range, superconducting - like proximity effect in bulk graphite outfitted with superconducting current leads. The phenomenon manifests as SC-like transitions on graphite's resistance, which was probed $700$ $\mu$m - $1$ mm away from the superconducting electrodes (much above $\xi_N\approx1-2$ $\mu$m). The suppression of these transitions is closely related to the breakdown of superconductivity on the current probes, thus suggesting that an unconventional superconducting proximity effect is at play. Our work supports the existence of intrinsic superconducting correlations in low-dimensional, localized regions of pristine graphite. Unfortunately, structural traits responsible for such properties could not be pinned from our measurements. However, results seem to discard features commonly found in the surface of graphite as their source. Our observations open routes towards superconducting electronic circuitry in bulk materials regardless of their volumetric conductivity.  

\section*{Acknowledgments}
We would like to thank Marta Cieplak, A. E. Koshelev, V. M. Vinokur, G. Baskaran and V. Khodel for fruitful discussions. This work was supported by the National Science Center, Poland, research project no. 2016/23/P/ST3/03514 within the The POLONEZ programme. The POLONEZ programme has received funding from the European Union's Horizon 2020 research and innovation programme under the Marie Sklodowska-Curie grant agreement No. 665778. P.G. acknowledges the support of the National Science Center, Poland, research project no. 2014/15/B/ST3/03889. Y.K. was supported by FAPESP, CNPq and AFOSR Grant FA9550-17-1-0132. The research of B.C., P. Sz. and M. K. performed in Slovakia was supported by European Microkelvin Platform (EU's H2020 project under grant agreement no. 824109) and projects VA SR ITMS2014+ 313011W856, APVV-18-0358, VEGA 1/0743/19.

%\section*{Supplementary information}
%The supplementary information contains sample pictures, raw data, measurements in additional samples, measurements in different systems, null results and the characterization of the %In/Pb alloy used in this work.

%\section*{References}


%\bibliography{Database}


\end{document}